\documentclass{aa}
\usepackage{graphicx}
\usepackage{natbib}
\usepackage{verbatim}
\usepackage{txfonts}
\usepackage{amsfonts}
\usepackage{amssymb}\bibpunct{(}{)}{;}{a}{}{,}

\def\bega{\beta_\Gamma}

\begin{document}
\title{The impact of thermal gas in AGN jets on the low-frequency emission}
\author{M. Siewert\inst{1} \and M. Pohl\inst{1,2} \and R. Schlickeiser\inst{1}}
\institute{Institut f\"ur Theoretische Physik, Lehrstuhl IV,
 Ruhr-Universit\"at Bochum, 44780 Bochum, Germany \and
Department of Physics and Astronomy, Iowa State University, Ames, IA 50011, USA}
\date{Received 29 March 2004 / Accepted 25 May 2004}

\abstract{
We study the effect of non-relativistic, thermal matter in the jets of
active galaxies (AGN)
on the low-frequency non-thermal emission and the variability thereof. In 
matter-dominated jets, sizable quantities of gas should exist, in particular
in the compression zones near the collision fronts that are an implicit ingredient of
Fermi-type particle acceleration scenarios. Non-relativistic thermal gas in AGN jets
noticably contributes to the optical depth at radio to infrared frequencies, and much less
to the emission, with an efficiency that is strongly temperature-dependent.
The observable flux of
low-frequency emission is thus modulated by the temperature evolution of the
thermal gas, and it can therefore display very complicated variability.
For a particular particle energisation scenario we calculate the temperature evolution of the thermal
plasma as well as the radiation transport of low-frequency emission, and thus derive simulated
light curves at different frequencies and their typical correlation properties.
}

\offprints{M. Siewert, e-mail:\tt ms@tp4.rub.de}

\maketitle

\section{Introduction}
Blazars, a sub-class of active galactic nuclei (AGN), are among
the strongest known radiation sources in the universe.
Observations show that they exhibit strong optical polarization,
variability on all observable timescales,
flat-spectrum radio emission from a compact core,
and many of them emit the bulk of their luminosity in the form of
$\gamma$-rays \citep[e.g.][]{der95obs,muk97}.
On account of these measurements
it is usually argued that the $\gamma$-rays are produced in
very compact regions within a relativistically moving system, the so-called
jets, thus providing a strong Doppler amplification of the radiation. 
Often, the relativistic bulk motion can be directly observed as apparent
superluminal motion of individual emission regions in the jets
in sequences of VLBI observations of their radio emission. 
The Lorentz factors (and Doppler factors)
thus derived are of the order of ten for general samples of AGNs
\citep{vc94}, but may be higher for AGNs showing
prominent gamma-ray emission \citep[e.g.][]{hom02,hom03}.
The range of Lorentz factors prevalent at the time of gamma-ray
emission, which presumably occurs before the emission region becomes visible
at radio frequencies, is not known, but is likely higher than ten, if 
the bulk kinetic energy of the jets is the energy reservoir for the particle
acceleration \citep[see, e.g., ][]{gk03}. These
conclusions are further supported by the absence of $\gamma-\gamma$ pair
absorption and the violation of the Elliot-Shapiro relation
\citep{es74relation} in several observed AGNs.

The unified model for active galaxies \citep{urry95} assumes that a
supermassive black hole is located in the center of a galaxy. The black
hole supposedly accelerates plasma to high energies, while the surrounding
host galaxy provides a steady inflow of matter, often assumed
to be in the form of an accretion disk,
to keep the system running for a long time.
Above the accretion
disk plasma is ejected in jets, which may reach
lengths of several hundreds kpc \citep{bbr84theory,rsch02book}.

Though thermal radiation can be observed from many AGN,
it is usually associated with
the accretion disk and the medium around it. The emission from 
the jets is generally assumed to be entirely non-thermal on account of the 
spectrum and the variability behaviour. Nevertheless, there may be
thermal plasma in the jets, in which the dominantly radiating,
energetic particles are embedded and confined.

In this paper we will study the effect of thermal matter in the jets
on the low-frequency non-thermal emission and the variability thereof.
We will do so in the framework of one particular model of particle 
acceleration, in which the energetic particles are provided by the
isotropization of interstellar matter in the downstream region of a
relativistic collision front \citep[][henceforth
referred to as PS]{ps00blast},
though most of our treatment, and hence most of the results, are
not restricted to the particulars of this scenario.

As we will see,
the thermal gas will not necessarily manifest itself by its emission,
but rather by its absorption. Both the emission and the absorption 
properties of thermal matter depend strongly on the temperature,
and thus the problem at hand has two aspects.
First, the radiation transport of low-frequency emission through 
thermal gas must be calculated, which is a function of the plasma temperature
and density as well as of the spectrum of energetic particles.
We then need to follow the temperature evolution of the thermal gas,
which depends on the its emission and absorptions coefficients
(and hence on its temperature), on the spectrum of energetic particles,
and on possible wave damping. In this study we concentrate on continuum emission
processes.

The spectrum of non-thermal particles 
obviously varies with time on account of the variability in the high-energy 
emission. Consequently the plasma temperature will vary, and so will
the thermal absorption coefficient. The observable flux of
low-frequency emission is thus modulated by the temperature evolution of the
thermal gas, and it
can therefore display very complicated variability
behaviour, the study of which is the subject of this paper.

We first give a brief introduction into the
PS model for active galactic nuclei
(Sect. \ref{chap-model}). After this, in Sect. \ref{chap-radtrans}
we discuss the radiation transport for a general non-thermal
population of electrons in a warm electron-proton plasma.
We present numerical absorption and emission coefficients
for a generic situation (when this is possible) as well as for
the PS model (when a generic treatment is not possible).
In Sect. \ref{chap-temptheo} we discuss the temperature evolution
in a generic warm background plasma for different heating
processes. The results of this analysis are independent of the non-thermal
particle spectrum, so that they can easily be applied to other
(non-)astrophysical situations. 

In Sect. \ref{chap-numerics} and \ref{chap-tempevol} we present numerical
results for the PS model and demonstrate that only a
few parameters actually influence the qualitative behaviour of
a system. Section \ref{chap-tempevol} emphasizes the limits of
our calculations, while Sect. \ref{chap-parameter} connects these
limits to free parameters in this situation.

\section{The basic model}
\label{chap-model}
\citet{ps00blast} have studied the kinetic
relaxation of particles that have traversed a parallel collision front
from the
upstream to the downstream region. This pick-up process occurs due to 
scattering off low-wavenumber plasma waves that the picked-up 
particles generate through streaming instabilities.
The calculation describes the first half-cycle of the standard shock
acceleration process for relativistic, collisionless flows.
If the outflow plasma propagating through the jet is dense,
the downstream region provides a target for the 
ultra-relativistic protons to interact with and produce radiation
through various leptonic and hadronic emission channels.
The predicted radiation properties resemble those of gamma-ray blazars
and thus the model offers an
attractive alternative to shock acceleration of electrons and subsequent
Inverse Compton scattering.

\begin{figure}
\begin{center}
\resizebox{0.8\hsize}{!}{\includegraphics{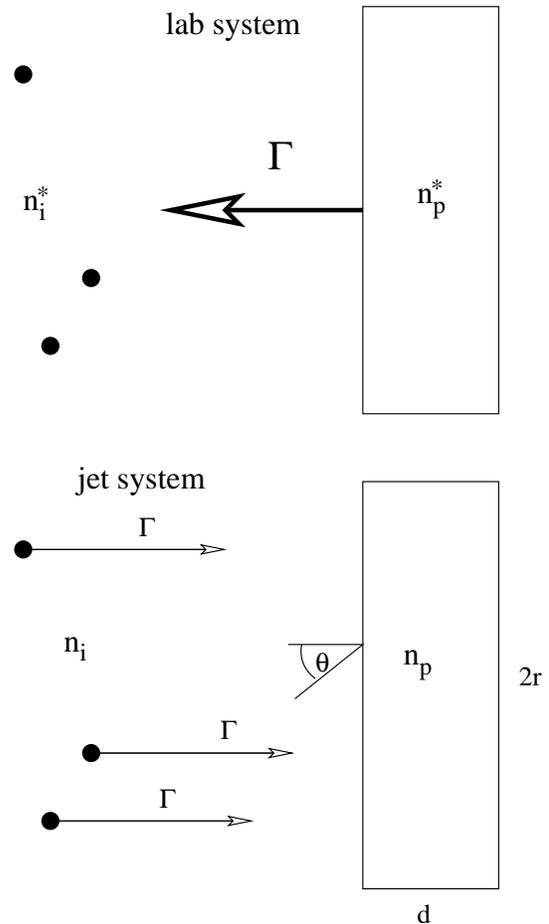}}
\end{center}
\caption{The geometry visualized for two systems of reference, the 
jet frame and the laboratory (host galaxy) frame, for which all quantities
are indexed with an asterisk. We consider a single cloud of
thermal proton-electron plasma with density $n_{\rm p}$, that is relativistically moving through
the interstellar medium of the host galaxy with density $n_{\rm i}^\ast$.
An observer would see radiation from the plasma cloud that is emitted at an
angle $\theta$ with respect to its direction of motion.}
\label{fig-pickup}
\end{figure}

The geometry of the system is shown in Fig.~\ref{fig-pickup} for two
systems of reference, the jet frame (without index)
and the laboratory (host galaxy) frame, for which all quantities
are indexed with an asterisk.
In this report, we also use a third reference frame, the observer frame,
which we use to calculate the observed evolution time. We will come back
to this point later in this section.
We consider the jet to consist of individual clouds of
thermal proton-electron plasma, that are relativistically moving through the interstellar
medium of the host galaxy. Here we discuss the interactions of one such cloud 
with temperature $T$, density $n_{\rm p}$, and as a simple
approximation for its spatial extent and form 
a cylindrical disk with radius $r$ and thickness $d$.
We emphasize that we treat the swept-up matter as a
simple collection of particles, which is in accordance with the earlier
model by \citet{ps00blast}, and unlike the fluid treatment common
to shock acceleration physics \citep{rsch02book}.
As a consequence, there is no relation between
the two quantities $n_p$ and $n_i$, and the
Rankine-Hugoniot conditions do not apply.

The Lorentz transformation between the two frames of reference gives the relations
$n_{\rm i} = \Gamma n_{\rm i}^\ast$ and $n_{\rm p}^\ast = \Gamma n_{\rm p}$
with a relative Lorentz factor of $\Gamma$.
The swept-up particles are of high
energies in the instantaneous downstream (jet) frame and are quickly isotropized
\citep{ps00blast}. Momentum conservation causes the system
to decelerate, while internal processes produce a broad distribution
of primary protons, $N_{\rm p} (\gamma_{\rm p})$, and secondary electrons,
$N_{\rm e} (\gamma_{\rm e})$, which are more important than primary electrons
for the parameters of interest.
Unless noted otherwise, all relevant expressions use cgs units,
all particle spectra $N_{\rm x}$ are taken per unit volume
and all equations and formulae are taken in the instantaneous
jet frame.

The isotropization of the swept-up particles effects a momentum transfer from
the interstellar medium particles to the jet. 
Therefore, the jet decelerates and the Lorentz factor of
the jet, $\Gamma$, is a function of time. It follows the equations
\citep{ps00blast}
\begin{eqnarray}
\dot{\Gamma} &=& -\frac{\pi r^2 n_i^\ast c m_{\rm p}}{M(t)}
  (\Gamma^2 - 1)^{3/2}
 \label{eq-Gammadot} \\
M(t) &=& V \sum_{\rm x=e,p} \int_1^\infty d\gamma\ 
 m_{\rm x}\,\left(n_b + \gamma \,N_{\rm x}(\gamma,t)\right)
\end{eqnarray}
where $M(t)$ is the total mass of the compact plasma, which in turn
depends on the differential number density of the energetic
particles, $N(\gamma)$. Such gradual deceleration has been directly 
observed in the case of the micro-quasar XTE~J1550-564 \citep{corbel02}.

The Lorentz factors observed in astrophysical sources vary greatly
depending on the source class and the measured velocity indicator. While in
gamma-ray burst (GRBs)
they may reach values of several hundreds, the Lorentz factors in
AGN jets are usually only of the order of ten, if deduced from measurements 
of superluminal motion, but may be higher for $\gamma$-ray-bright blazars. 
For TeV-blazars the radiation modeling often also requires significantly 
higher Lorentz (and Doppler) factors \citep[e.g.][]{kono03}. However,
gamma-ray emission is presumed
to occur before the emission zone becomes visible at radio frequencies,
and the corresponding Lorentz factor should thus be very much higher than
ten, if the bulk kinetic energy of the jets is supposed to be
the source for particle acceleration. Additionally, the physics
of jet formation, and therefore the Lorentz factors involved
with this process are not understood, but considering the supposedly
extreme conditions near a black hole, it is highly possible
that the particles which are ultimately injected into the jet
initially possess similarly high velocities. For this reason, we assume
an initial Lorentz factor $\Gamma_0=300$. We would like to point out that
our results are only parametrically dependent on this choice.

In the instantaneous jet frame the differential number density of protons and electrons
follows the evolution equation
\begin{equation}
\frac{\partial N}{\partial t}=\dot{N}-\frac{\partial}{\partial \gamma}
 \left( \dot{\gamma}N(\gamma) \right)-\frac{N(\gamma)}{T_{\rm c}}
\label{eq-evol}
\end{equation}
where $T_{\rm c}$ is the timescale for particle losses
by escape or decay,
$\dot{\gamma}$ represents continuous energy
losses, and
$\dot{N}$ is the injection or production rate of new particles.
As initial condition at the time $t=0$, when the jet Lorentz factor is
$\Gamma=\Gamma_0$, we assume that there are no energetic particles in the jet,
$N(\gamma)=0$.

In the default configuration the background plasma sweeps
up protons from the interstellar matter, while
electrons are produced in inelastic collisions in the
system.
The differential proton source density in the co-moving frame is
\citep{ps00blast}
\begin{eqnarray}
\dot{N}= {{c\, n_{\rm i}^\ast}\over d} \sqrt{\Gamma^2-1} \; \delta(\gamma-\Gamma)
\label{eq-inje}
\end{eqnarray}
while the (secondary) electron production proceeds through
the reaction chain
\begin{eqnarray}
p+p &\rightarrow& \pi^\pm + X \\
\pi^\pm &\rightarrow& \mu^\pm + \nu \\
\mu^\pm &\rightarrow& e^\pm + 2\nu,
\end{eqnarray}
and is calculated numerically to properly account for the pion multiplicity
spectrum in reaction (3).
Here we use the Monte Carlo model DTUNUC (V2.2)
\citep{mr91dtunuc,ranft94dtunuc,ferr96dtunuc,dtunuc97},
which is based on a dual parton model \citep{cap94dualparton}.
This MC model for hadron-nucleus and nucleus-nucleus interactions
includes various modern aspects of high-energy physics and has been
successfully applied to the description of hadron production in
high-energy collisions \citep{Fer96b,RR94,Moe93,RER98}.

The particle losses are diffusive escape (protons and
electrons), neutron escape following $\hbox{p+p}\rightarrow \hbox{n+X}$ reactions
and pair annihilation (electrons and positrons),
\begin{equation}
T_{\rm esc} = 1.67\cdot 10^{-19}\ \frac{d^2\,\Gamma\, n_i^\ast}{\beta \sqrt{n_{\rm p}}}
\quad \hbox{sec}
\label{eq-tesc} 
\end{equation}
\begin{equation}
T_{\rm neu} = 3\cdot 10^{15} n_{\rm p}^{-1} \left< \exp \left(-
 \frac{L}{\gamma\, 2.7\cdot 10^{13}\ {\rm cm}} \right) \right>^{-1}\quad \hbox{sec}
\label{eq-tneut}
\end{equation}
\begin{equation}
T_{\rm ann}^{-1} = \frac{3c\sigma_T n_{\rm p}}{8(\gamma+1)\sqrt{\gamma^2-1}}\times
\label{eq-tann} 
\end{equation} 
$$ \left[(\gamma+4+\gamma^{-1})\log(\gamma+\sqrt{\gamma^2-1})-\beta(\gamma+3)
  \right]\quad \hbox{sec}^{-1} \nonumber $$ 
\begin{equation}
\frac{1}{T_{\rm c,tot}} = \sum_i \frac{1}{T_{\hbox{c,}i}}.
\end{equation}
For all relevant cases the exponential in eq. \ref{eq-tneut} (where $L$
is the distance to the plasma cloud boundary)
can be approximated by 1 \citep{ps00blast}.
We approximate the continuous losses by
\begin{eqnarray}
-\dot{\gamma}_{\rm el,p}&=&3.75\cdot10^{-16}\ n_{\rm p}\,\frac{\gamma}
 {\sqrt{\gamma^2-1}}\quad \hbox{sec}^{-1} \label{eq-gdotelp} \\
-\dot{\gamma}_{\rm el,e}&=&7.5\cdot10^{-13}\ n_{\rm p}\,\frac{\gamma}
 {\sqrt{\gamma^2-1}}\quad \hbox{sec}^{-1} \label{eq-gdotele} \\
-\dot{\gamma}_{\rm inel,p}&=&7\cdot10^{-16}\ n_{\rm p}\,
        \frac{(\gamma-1)^2}{\gamma+1}\quad \hbox{sec}^{-1} \label{eq-gdotinelp} \\
-\dot{\gamma}_{\rm Sync}&=&1.1\cdot10^{-15}\ \frac{B^2}{m_{\rm e}c^2}
        \,\gamma^2\,\beta^2 \quad \hbox{sec}^{-1}\label{eq-gdotsync} \\
-\dot{\gamma}_{\rm Brems}&=&8\cdot10^{-16}\ n_{\rm p}\,(\gamma-\gamma^{-1})
 \quad \hbox{sec}^{-1}.
\label{eq-gdotbrems}
\end{eqnarray}
Here the subscripts $el$ and $inel$ stand for elastic and inelastic
processes, $Sync$ for
synchrotron radiation and $Brems$ for Bremsstrahlung. The
equations \ref{eq-gdotelp} and \ref{eq-gdotele} are taken
from \citet{h88coulomb}, eq. 18, assuming
a Coulomb logarithm of 25 and neglecting the last terms.
This modification has been made
to be consistent with eq. \ref{eq-heatcou}, while in
\citet{ps00blast} a value of 20 is used. It does not affect
the overall results in any visible way.
Equation \ref{eq-gdotsync} is from \citet{p80book},
averaged over all emission directions.
The inelastic energy losses,
eq. \ref{eq-gdotinelp} are a result of the model
used to evaluate these processes \citep{cap94dualparton}.
Finally, the Bremsstrahlung energy losses are taken from
\citet{haya69} and \citet{ps00blast}.

The timescales for particle losses are from \citet{jr76book} (eq. \ref{eq-tann})
and \citet{ps00blast} (eq. \ref{eq-tesc} and \ref{eq-tneut}).
The diffusive escape timescale used here is derived for the
disk geometry under the condition $\hbox{r}\gg \hbox{d}$, which implies that effectively
the charged particles will only escape through the top and
bottom surfaces. The most important point
here is the assumption that the energetic particles start diffusing outwards 
near the center of the jet plasma cloud. Depending on the Alfv\'en speed
and the initial intensity of scattering waves this may be questionable
\citep{vps03}.
We also want to emphasize that these equations have been
derived under the assumption that the plasma is fully ionized
and that the particle number is conserved.
If this is not the case, the above equations have to be modified
for ionization, recombination and pair production.

The free parameters of the model are specified in table
\ref{tab-params} for a quick reference, where we also show the
numerical values we have adopted as standard set of parameters in this paper.
Whenever we modify one of these values, it is explicitely mentioned
in the text.
\begin{table}
\begin{center}
\begin{tabular}{|l|c|}
\hline
parameter & default value \\
\hline
\hline
average magnetic field $B$ & 1 G \\
background plasma density $n_{\rm p}$ & $5\cdot 10^8 \hbox{cm}^{-3}$ \\
interstellar matter density $n_{\rm i}^\ast$ & $0.4\;\hbox{cm}^{-3}$ \\
observer angle $\theta^{\rm obs}$ & $0.1^\circ$ \\
initial Lorentz factor $\Gamma_0$ & 300 \\
disk height $d$ & $3\cdot 10^{13}\;\hbox{cm}$ \\
disk radius $r$ & $10^{14}\;\hbox{cm}$ \\
initial temperature $\rm T_0$ & $10^4 {\rm K}$ \\
\hline
\end{tabular}
\end{center}
\caption{Standard parameters used in this report unless noted
otherwise. For the relation between the observer angle
and the emission angle see Eq.\ref{eq-theta2obs} and \ref{eq-thetafromobs}.
}
\label{tab-params}
\end{table}
We have neglected all $e^+e^-$ annihilation processes,
since we focus on the optical-to-infrared region where
pair annihilation does not contribute.

All calculations in this analysis are made in the
jet rest frame. For a comparison with observations, the
photon spectra have to be transformed into the observer's frame,
which, depending on the cosmological redshift of the AGN in question,
can be different from the host galaxy frame. For ease of exposition 
we will assume in the following treatment that the observer's
frame and the host galaxy frame are identical, as was done in
\citet{ps00blast}.

As a reminder we here list the relations between relevant
quantities in the observer's frame and the jet frame.
While the observer's frame is fixed, the jet frame is only the 
instantaneous rest frame of the jet plasma at the time considered.
Since we are interested in the long-time evolution
of a compact relativistic object moving along a
straight line under a \emph{fixed} aspect angle in the 
observer's frame, we also need to
consider that the corresponding aspect angle in the instantaneous jet 
frame is not constant, for the Doppler factor is changing with time.
\begin{eqnarray}
\cos\theta^{obs} &=& \frac{\cos\theta + \bega}{1+\bega\cos\theta}
 \label{eq-theta2obs} \\
\cos\theta &=& \frac{\cos\theta^{obs} - \bega}{1-\bega\cos\theta^{obs}}
  \label{eq-thetafromobs} \\
D &=& \frac{1}{\Gamma(1-\bega \cos\theta^{obs})} \label{eq-doppler} \\
\nu^{obs} &=& D\nu \\
I^{obs}(\nu^{obs}) &=& D^3 I(\nu^{obs}/D) \\
dt^{obs} &=& dt/D
\end{eqnarray}

\section{The radiation transport}
\label{chap-radtrans}
\subsection{The radiation transport equation}
To calculate the emitted photon spectra $I_{\nu,0}$ for a given line-of-sight,
it is required to know
the emission coefficient $j_\nu$ and the absorption coefficient
$\alpha_\nu$. These two quantities enter in the radiation transport
equation \citep[e. g.][]{rl85radiation},
\begin{equation}
\frac{dI_{\nu,0}}{d\tau}=S_\nu - I_{\nu,0}
\label{eq-radtrans}
\end{equation}
via the source function $S_\nu = \sum_i j_i / \sum_i \alpha_i$
and the optical depth
\begin{equation}
\tau=\int_{0}^{s}\alpha\;dx\ .
\label{eq-thick}
\end{equation}
The well-known general solution to this equation are for a path length
of $s$
\begin{equation}
\begin{array}{llr}
I_{\nu,0}(\tau)&= S_\nu\,\left[1-\exp(-\tau)\right] \label{eq-inu1}& \\
&=j_\nu\cdot s &\quad\hbox{for $\tau \ll$ 1}\hphantom{.} \label{eq-inu1a} \\
&=S_\nu &\quad\hbox{for $\tau \gg$ 1}.
\label{eq-inu1b}
\end{array}
\end{equation}
In our calculations we have assumed that the
radiation coefficients $j_\nu$ and $\alpha_\nu$ are independent of location,
i.e. the supposedly cylindrical cloud of plasma in the jet is homogeneous,
so that the optical depth reduces to
\begin{equation}
\tau_s=\alpha\cdot s
\label{eq-taux}
\end{equation}

These expressions are the solution of the radiation transport equation
for a single line-of-sight through the jet plasma.
We still have to integrate
over the entire emitting surface, as well as respecting the exact
geometry of the system, since the photon path length is
not constant over the apparent surface of the emission region.
The total spectral power is then
\begin{eqnarray}
P_\nu (\Omega) &=& \int dA(\Omega)  \ I_{\nu,0} (\Omega)
\label{intens}
\end{eqnarray}
with the apparent surface element $dA$ and the emission direction
$\Omega$.

Since a real cloud of plasma will have a complicated (and generally unknown)
geometry rather than the simple shape considered
by \citet{ps00blast} and in this report,
it may be sufficient to arrive at an approximate
solution to Eq.\ref{intens}.

Let us start with the most simple approximation: we assume
that the path length, $s$, is independent of
the emission point both for the front surface, $A_{\rm f}$, and the side 
surface, $A_{\rm s}$. Then for the front surface
\begin{eqnarray}
s_f &\simeq& d/\cos\theta,
\end{eqnarray}
and
\begin{eqnarray}
P_{\nu,f}(\Omega) &\simeq& A_f \; \cos\theta \; I_{\nu,0}(d\theta) \\
 &\simeq&\pi\,r^2\,\cos\theta\, S_\nu\,\left[1-\exp\left(-{{\alpha\,d}\over {\cos\theta}}\right)\right].
\label{eq-emisstheta}
\end{eqnarray}
On account of the assumed thin-disk geometry, i.e. $d\ll r$,
the contribution from the
side surface is significant only for emission angles $\theta\approx \pi/2$.
The choice of a thin disk for our system
introduces an additional problem, for the photon path length
between two opposite points on the side surface is not constant; 
we use $s \simeq r$ and obtain
\begin{eqnarray}
s_s &\approx& r/\sin\theta,
\end{eqnarray}
and
\begin{eqnarray}
P_{\nu,s}(\Omega) &=& A_s \; \sin\theta \; I_{\nu,0}(d\theta) \\
 &=&2\,d\,r\,\sin\theta\, S_\nu\,\left[1-\exp\left(-{{\alpha\,r}\over {\sin\theta}}\right)\right].
\label{eq-emisstheta2}
\end{eqnarray}
which turns out to be a fair approximation to the exact
solutions of the problem, which
are derived in appendix \ref{app-geometry}.

A better approximation is afforded by a box geometry instead of the
original cylinder.
The side lengths would be $a=\sqrt{\pi}\,r$ to conserve the volume.
If one assumes that
the line-of-sight to the observer is in the plane of two of the side
surfaces, Eq.\ref{intens}
can be solved exactly. The resulting expressions
for the observable photon spectra are presented in appendix
\ref{app-box}. The box geometry turns out
an excellent approximation to the exact solution for the initially assumed
disk geometry, with a maximum error of 3\% at $\theta=60^\circ$,
while the simple constant-$s$ approach introduces an error of 20-30\%.

\subsection{The emission coefficients}
In this report we consider two different contributions to the radiation
processes, synchrotron emission and thermal bremsstrahlung
(free-free emission). The bremsstrahlung coefficients for a
quasineutral thermal hydrogen plasma are
\citep{rl85radiation}
\begin{equation}
j_{br} =\frac{\sqrt{2}\,n^2\,\alpha\,\sigma_T\,\hbar\, c}{\pi^{3/2}}
\sqrt{\frac{mc^2}{kT}}\,\ln\left({{2\,kT}\over {h\nu}}\right)\,
\exp\left(-\frac{h\nu}{kT}\right)
\label{eq-jth} 
\end{equation}
\begin{displaymath}
\alpha_{br} = \frac{2\,n^2\,\alpha\,\sigma_T\, c^3}{(2\pi)^{5/2}\,\nu^3}
\sqrt{\frac{mc^2}{kT}}\,\ln\left({{2\,kT}\over {h\nu}}\right)
\left[1-\exp\left(-\frac{h\nu}{kT}\right)\right]\, .
\end{displaymath}
\begin{equation}
\ 
\label{eq-alphath}
\end{equation}
\begin{figure}
\resizebox{\hsize}{!}{\includegraphics{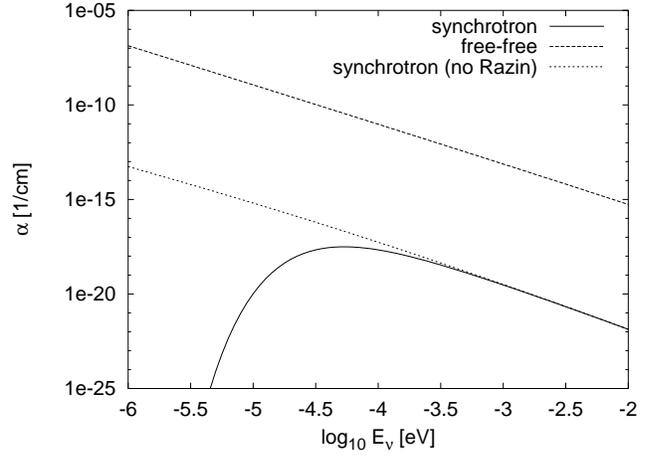}}
\caption{The absorption coefficients $\alpha$ for the free-free process
and the synchrotron process for the jet model of \citet{ps00blast}
with standard parameters (see table \ref{tab-params}).
We display the situation
after one hour in the observer's frame, at which time the plasma 
temperature is T$\approx 3\cdot 10^4\,$K.
Apparently the system is dominated by
free-free absorption.}
\label{fig-alphadef}
\end{figure}

\begin{figure}
\resizebox{\hsize}{!}{\includegraphics{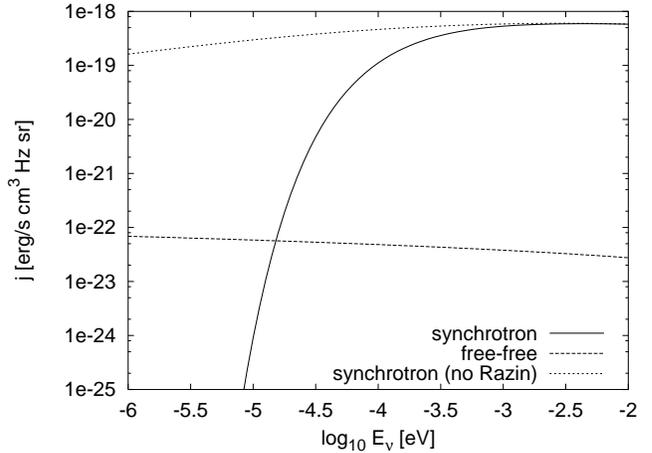}}
\caption{The emission coefficients $j$ for the free-free process
and the synchrotron process after one hour observed time for
standard parameters, as in Fig.\ref{fig-alphadef}.
The free-free emission is only
visible because of the Razin effect.
For $n_{\rm p} = 5\cdot 10^8\ {\rm cm}^{-3}$ the plasma frequency corresponds
to $E_\nu \approx 8 \cdot 10^{-7}\ {\rm eV}$, so we cover
the entire valid energy spectrum.
}
\label{fig-jdef}
\end{figure}

The synchrotron radiation coefficients for an ensemble of electrons with 
differential number density $N(\gamma)$ can be calculated as
\citep{rl85radiation}
\begin{eqnarray}
j_{\rm sy} &=& \frac{1}{4\pi}\int_1^\infty d\gamma\; N(\gamma)\,p_{sy}(\nu)
\label{eq-jsync} \\
\alpha_{\rm sy} &=& -\frac{1}{8\pi m \nu^2}\int_1^\infty d\gamma\;
  (\gamma^2-1)\,p_{sy}(\nu)\, \frac{\partial}{\partial \gamma}
  \left(\frac{N(\gamma)}{\gamma^2-1}\right)
\label{eq-alphasync}
\end{eqnarray}
with the spectral power per electron
\begin{equation}
p_{sy}(\nu) = \frac{\sqrt{3}e^3B}{mc^2}\,F \left( \frac{\nu}{\nu_c} \right)
\end{equation}
where
\begin{equation}
\nu_c = \frac{3eB}{4\pi mc}\gamma^2
\qquad\qquad
F(x) \approx 1.8\; x^{1\over 3}\,\exp(-x)
\label{eq-fvonx}
\end{equation}
In a dense plasma with plasma frequency $\nu_{\rm p}$ the
spectral power is modified at low frequencies by the Razin effect
\citep{rl85radiation,cs88rad} on account of the modified dispersion
relation for electromagnetic waves. Then
\begin{equation}
p_{sy,Razin}(\nu) = \frac{p_{sy}(\nu R^{3/2})}{\sqrt{R}}
\end{equation}
where
\begin{equation}
R = 1+\left( \frac{\nu_{\rm p}\gamma}{\nu} \right)^2 \qquad\qquad
\nu_{\rm p} \simeq 9\cdot 10^3 \sqrt{n_{\rm p}}\ {\rm Hz}.
\label{eq-razi}
\end{equation}
The Razin effect creates a cutoff in the photon spectrum at low energies,
whereas at high energies the emission spectrum is unchanged.

The radiation coefficients for the standard parameters 
(see table \ref{tab-params}) in the jet model of \citet{ps00blast} are
displayed in Fig.\ref{fig-alphadef} and Fig.\ref{fig-jdef}.
The synchrotron coefficients are based on the numerically derived
electron spectrum in the system after one hour of observed time, while the
free-free radiation coefficients depend only on the temperature,
which is approximately $3\cdot 10^4\ {\rm K}$ at this time. The
kinematic of the pion and muon decays is unfavorable for the production of 
secondary electrons with an energy less than approximately 100~MeV. Thus,
essentially only highly relativistic electron are generated. The
energy loss rate below 300~MeV is dominated
by Coulomb and ionization interactions, i.e. is independent of energy,
so that in a steady-state spectrum the 
cooling tail towards lower energies would be flat. The synchrotron
spectrum in Fig.~\ref{fig-jdef} is correspondingly inverted.

\begin{figure}
\resizebox{\hsize}{!}{\includegraphics{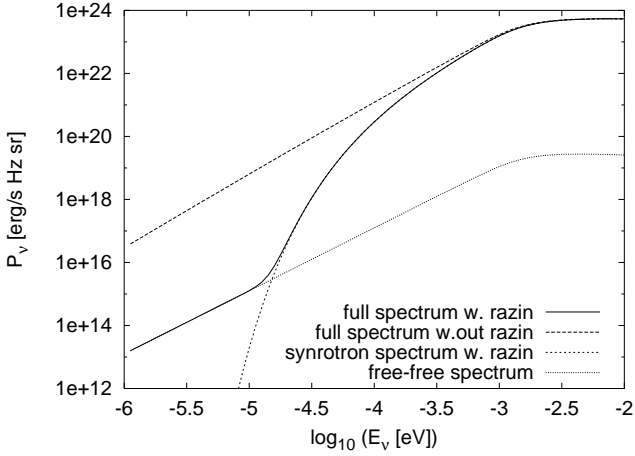}}
\caption{Emitted photon spectra after one hour of observed time
in the jet system for the standard parameters
($\theta^{\rm obs}\approx 0.1^\circ$, see table \ref{tab-params}) and
$T \approx 3\cdot 10^4\,$K.
The thermal Bremsstrahlung dominates at low energies only because
of the Razin-effect, which suppresses the synchrotron spectrum here.}
\label{fig-spectra}
\end{figure}

%
%
\begin{figure*}[t]
\centering
\includegraphics[width=\textwidth]{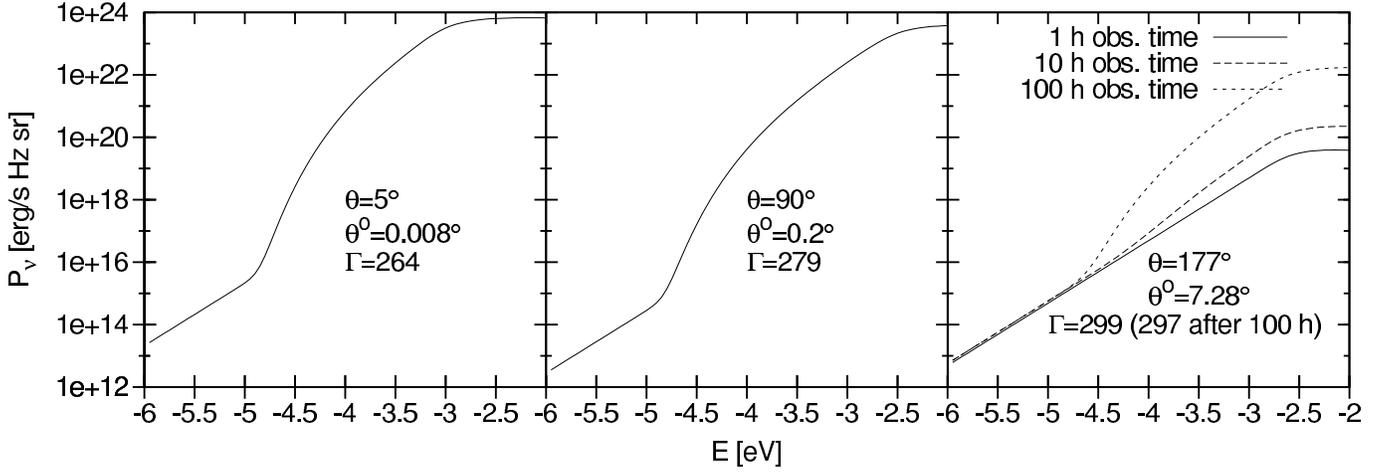}
\caption{Emitted photon spectra in the jet frame
after one hour of observed time
for different emission angles, {based on an exact treatment of the geometry}.
For high $\theta^{\rm obs}$, the slow evolution of the synchrotron part of
the spectrum has been reproduced.}
\label{fig-winkel}
\end{figure*}
 
The non-thermal emission
spectra do not significantly change when we modify the parameters,
except for an approximately linear dependence on the density
of the interstellar medium,
so the results displayed here are a good estimate in most situations.
The free-free absorption, and to a smaller extent the emission, is strongly
reduced in efficiency for
higher temperatures. As we shall see later, the plasma temperature can rise from
$T=3\cdot 10^4\,$K, which nearly is a lower limit, to $10^9\,$K or more.
Modifying the temperature from $3\cdot 10^4\,$K
to this value reduces the absorption
coefficient by 7 orders of magnitude, which makes the free-free
absorption less efficient than the synchrotron absorption. The temperature,
for which both are approximately equal, is around $10^8\,$K.

It thus appears that the optical thickness is mostly dominated by
free-free absorption, while the emission is dominated by
either synchrotron radiation or bremsstrahlung, depending on the frequency.
It is important to include the Razin effect in our calculations,
for, without the Razin effect, the
synchrotron emission coefficient would not have a cut-off at low frequencies,
and free-free emission would be completely negligible.

An interesting possibility is that a synchrotron maser may operate
in the jet plasma.
Maser activity requires a negative absorption coefficient
($\alpha < 0$), indicating that stimulated emission is
more important than the spontaneous absorption. If the optical depth
$\tau \ll -1$, the emitted photon intensity can be very high.

\citet{cs88rad} have shown that for the synchrotron process this
is only possible when the electron spectrum
has a slope of at least 2, i.e. the distribution function is inverted.
The numerical calculations of the inelastic proton decay chain
seem to indicate that this is not the case, at least in the
energy regions covered in this report. The condition for
synchrotron masering is $\alpha < 0$, and since the evaluation of $\alpha$
requires an integration over the entire radiating particle distribution
(eq. \ref{eq-alphasync}), this condition must be valid for a large energy
interval, and a \emph{locally} hard spectrum will not be sufficient
for a maser to operate.

\section{Simple photon spectra}
\label{chap-numerics}
In this section we discuss radiation spectra 
for a constant temperature with a view to understand
the characteristic properties of the escaping 
low-frequency emission. After one hour of observed time, the
temperature is $T \approx 3\cdot 10^4\ {\rm K} $.

Photon spectra calculated on the basis of the exact disk geometry
are presented in Fig. \ref{fig-spectra}.
The spectrum can be separated in an optically thick region
and an optically thin region. The optical depth itself is
dominated by the free-free absorption process.
The emission process is dominated by the synchrotron process, for which
the Razin effect causes a drop off below a certain
photon energy, so that at very low frequencies the free-free emission
becomes visible.

The spectra shown here are calculated in the jet frame, 
and thus they need to be 
transformed in the observer system for a comparison with data.
For blazars, this would result in a frequency shift by about
two orders of magnitude, so the turn-over frequency, at which the emission
transitions from optically thick to optically thin conditions,
would be located in the near-infrared.

Next, we calculate the emitted photon spectra for different
observation angles $\theta^{\rm obs}$ and show typical results in Fig.\ref{fig-winkel}.
For very small aspect angles $\theta^{\rm obs} \leq 1^\circ$, the jet system appears
to rapidly evolve;
in fact $\Gamma \approx 270$ after one hour of observed time.
This value is still very high compared to typical
Lorentz factors of 10, but one should note that it takes months for an
outburst to become visible at radio frequencies, where most
measurement of jet velocity are made.

The characteristic aspect of these spectra is the presence of
both the thermal and the synchrotron components of the spectrum.
Variations on the high-frequency part of the spectrum, most notably a shift
of the turn-over frequency, are visible, which result from the different
aspect in the jet frame, i.e. the different optical depth
for a frontal and side view of the emission region in the jet frame.

For larger observation angles ($\theta^{obs} > 1^\circ$), the
system evolves much slower, so that $\Gamma \approx \Gamma_0$
after a single hour.

\section{The evolution of the temperature}
Throughout this paper we have assumed that LTE is valid, that the jet plasma
is always non-relativistic, and that the
electron and proton components have the same temperature, i.e. the internal
equilibration processes are faster than the energy exchange with the
environment.
\label{chap-temptheo}
\subsection{The cooling process}
\begin{figure}
\resizebox{\hsize}{!}{\includegraphics{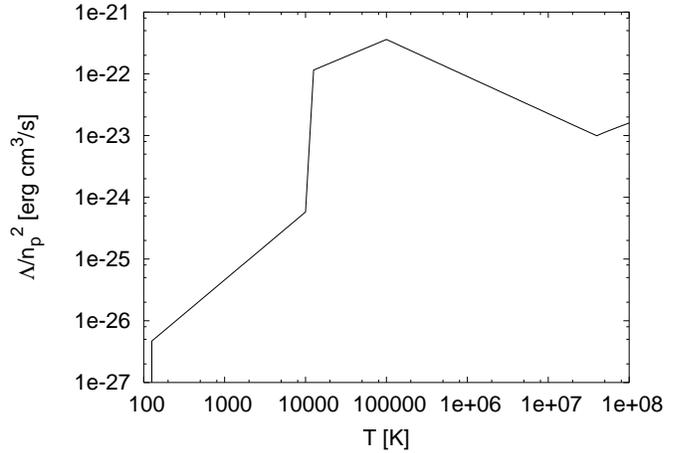}}
\caption{The approximated cooling function according to
Eq.\ref{eq-lambda}.}
\label{fig-lambda}
\end{figure}

As we have seen, the temperature of the jet plasma is a decisive parameter
for the radiation yield at low frequencies. The plasma is subjected to 
a number of heating and cooling processes, and therefore the temperature will
not remain constant. We use a balance equation to follow
the variations of the plasma temperature,
\begin{equation}
\dot{T}=\frac{1}{n_{\rm p} k}(\gamma-1)(\dot{\epsilon}-\Lambda)\ ,
\label{eq-tdot}
\end{equation}
with the adiabatic index $\gamma$ (=5/3 for a non-relativistic hydrogen gas),
the heating rate, $\dot{\epsilon}$, and the cooling rate, $\Lambda$.
For the cooling rate we use an analytical
approximation (see fig. \ref{fig-lambda}) of the standard interstellar
cooling function \citep{dmc72cool,rcs76cool}. 
\begin{displaymath}
\Lambda=\left(10^{-22}\ \hbox{erg cm}^{-3}\hbox{ s}^{-1}\right)\, n_{\rm p}^2
\end{displaymath}
\begin{eqnarray}
\hphantom{\Lambda=}\times\,\left\{
  \begin{array}{rll}
  2.3\cdot10^{-5}\,&T^{1.1}&\mbox{for }\ 10^2\le T\le 10^4\\
  0.64\,&T^{0.55}&\mbox{for }\ 10^4\le T\le 10^5\\
  3.6\cdot10^{5}\,&T^{-0.6}\quad &\mbox{for }\ 10^5\le T\le 4\cdot10^7\\
  1.6\cdot10^{-3}\,&T^{0.5}&\mbox{for }\ T\ge 4\cdot10^7
  \end{array} \right.
\label{eq-lambda}
\end{eqnarray}
where the temperature, $T$, is in units of Kelvin.
The error introduced by this approximation is not significant,
for other uncertainities such as the real geometry
have a much greater impact on the final results. A possible exception
is the jump at $T\simeq 10^4\ {\rm K}$ that results from electron-impact
excitation and ionization of 
\ion{H}{i}, but may be modified here on account of the collisional
excitation and ionization by the energetic particles.

\subsection{The Coulomb heating process}
\label{chap-heat-el}
The heating rate is more difficult to calculate,
since there are several concurring contributions.
For a completely ionized plasma,
the heating rate from elastic Coulomb scattering
of a distribution of suprathermal electrons or protons, $N(\gamma)$, 
in a thermal plasma of temperature $T$ is given by \citep{bb62,h88coulomb}
\begin{equation}
\dot{\epsilon} = \int_1^{\infty} d\gamma \; N(\gamma)\,
 \frac{3}{2\beta}\,
 \sigma_T\, m_{\rm e} c^3\, n_{\rm p}\, \ln{\Lambda}
\label{eq-heatcou}
\end{equation}
where $\ln\Lambda \approx 25$ is the Coulomb logarithm.
This expression is valid for protons and electrons
(this follows from Eq.\ref{eq-gdotele} and \ref{eq-gdotelp}), so that usually
the non-thermal particles with the greatest number density
dominate this heating process. In most practical situations
in our model, this is the secondary-electron population.

If only relativistic particles are considered, $\gamma \gg 1$ and
$\beta\simeq 1$; then the heating rate scales linearly with the
total number density of high energy particles, $N_{\rm tot}$.
\begin{eqnarray}
\dot{\epsilon} &\simeq& \frac{3}{2}\, \sigma_T\, m_{\rm e} c^3\, n_{\rm p}\, N_{\rm tot}\, \ln{\Lambda}
\label{eq-heatcousimp}
\end{eqnarray}

\subsection{The synchrotron heating process}
\label{chap-heat-sync}
Heating is also effected by absorption
of non-thermal emission. In our case, the dominant process is
free-free absorption of synchrotron photons.
The synchrotron absorption coefficient is usually small compared to the
free-free absorption coefficient, and hence the emitted radiation is
almost exclusively absorbed by the free-free process.
The heating rate can be calculated as an integral over the deficit
in total luminosity caused by the absorption.
\begin{equation}
\dot{\epsilon} = \frac{1}{V} \int_0^\infty d\nu
 \left( 4\pi V j_\nu - 2\pi \int d\theta \ P_\nu (\Omega) \right)
\label{eq-photheat}
\end{equation}
The spectral power, $P_\nu (\Omega)$, is given by Eq.\ref{intens}
and strongly dependent on the emission angle, thus necessitating
the integral over $\theta$. In appendix \ref{app-geometry} we will discuss ways to arrive at
approximate solutions to $P_\nu(\Omega)$ with various degrees 
of accuracy.

\subsection{Comparing both heating processes}
\label{chap-heat-compare}
It may be useful to identify parameters, for which Coulomb heating
is always more efficient than heating by absorption, in which case 
solving Eq.\ref{eq-photheat} would not be necessary. For that purpose,
we neglect the Razin effect, for the very low-frequency
photons do not significantly contribute to the total heating rate.
We also note that the optical depth is strongly dependent on frequency,
thus effectively dividing the spectrum at the turn-over frequency $\nu_+$,
for which $\tau = 1$, in a low-frequency part at $\nu \le \nu_+$,
in which essentially all photons are absorbed,
and a high-frequency part at $\nu \ge \nu_+$, in which essentially
all photons escape.

We have noted before that free-free absorption is generally more efficient than
synchrotron absorption. The turn-over frequency $\nu_+$ is determined by the 
absorption coefficient (Eq.\ref{eq-alphath}) and the path length $s$. 
Setting the logarithmic term to the constant 
value $L$ we find in the limit $h\nu \ll kT$
\begin{eqnarray}
\nu_+&=&n_{\rm p}\,\sqrt{\frac{2\,\alpha\,\sigma_T\, \hbar\,c}{(2\pi)^{3/2}\,m_{\rm e}}}\,
\left(\frac{mc^2}{kT}\right)^{3/4}\,\sqrt{s}\,\sqrt{L}
\nonumber \\
&\approx&(10^{11}\,\hbox{\rm Hz})\left(\frac{n_{\rm p}}{\rm 10^9\,cm^{-3}}\right)
   \sqrt{\frac{s}{\rm 10^{14}\,cm}}\sqrt{\frac{L}{\rm 10}}
   \left(\frac{T}{\rm 10^6\,K}\right)^{-{3\over 4}}\nonumber\\
 & & \label{eq-nuplus}
\end{eqnarray}
With these approximations eq. \ref{eq-photheat} reduces
to a much simpler form. Using eq. \ref{eq-jsync}
and changing the order of integration we obtain
\begin{eqnarray}
\dot{\epsilon} &\leq& \int_0^{\nu_+} d\nu \; 4\pi j_\nu \\
&=& \int_1^\infty d\gamma\; N(\gamma) \int_0^{\nu_+} d\nu \;  p_{sy}(\nu)
=\int_1^\infty d\gamma\; N(\gamma) \,f(\gamma)_{sy}  .
\label{eq-heatsynceasy1}
\end{eqnarray}
This integral has the same mathematical form as that in Eq.\ref{eq-heatcou}
and Eq.\ref{eq-heatcousimp}, implying that is may be sufficient to compare the
weight functions for the heating processes, $f$.

For Coulomb heating by relativistic particles, the weight function, $f_{\rm el,e}$,
follows directly from eq. \ref{eq-heatcousimp},
\begin{eqnarray}
f(\gamma)_{\rm el,e} &\simeq& 6\cdot 10^{-19}\, n_{\rm p}\quad {\rm erg\,cm^3\,s^{-1}}\,.
\label{eq-wgcoulomb}
\end{eqnarray}
Since electrons and protons have the same weight functions,
$f(\gamma)_{\rm el,e}=f(\gamma)_{el,p}$, this
applies to protons as well,
which are usually less numerous that electrons and
therefore won't dominate the overall heating process.

The synchrotron weight function is more difficult, for
the integral over $\nu$ in Eq.\ref{eq-heatsynceasy1},
\begin{eqnarray}
f(\gamma)_{\rm sy} &=&
   \int_0^{\nu_+} d\nu \; p_{sy}(\nu) \nonumber \\
 &=& 1.8\,\frac{\sqrt{3}e^3}{m c^2}\, B
   \int_0^{\nu_+} d\nu\ \left( \frac{\nu}{\nu_c} \right)^{1/3}
  \exp(-\frac{\nu}{\nu_c}) 
\label{eq-fsync}
\end{eqnarray}
requires a numerical treatment for general $\nu_+$.
However, an approximate solution can be derived. For that purpose
we compare the electron energy loss rates for synchrotron emission
(Eq.\ref{eq-gdotsync})
and Coulomb scattering (Eq.\ref{eq-gdotele}). For electron Lorentz factors,
$\gamma$, smaller than the critical value
\begin{eqnarray}
\gamma_- &=& \max(1,g(n_p,B))\\
 g(n_p,B) &=& \sqrt{682 \frac{n}{B^2} m_{\rm e}c^2}
 \approx 0.024 \left( \frac{n_{\rm p}}{\rm cm^3} \right)^{1/2}
  \left( \frac{B}{\rm G} \right)^{-1}\ ,
\label{eq-gammaminus}
\end{eqnarray}
the electrons lose more energy by Coulomb scattering than by synchrotron
radiation. This implies that Coulomb interactions are the dominant
heating process, whatever the optical depth of the jet plasma.
Only at Lorentz factors $\gamma\ge\gamma_-$ will the electrons dump
a significant fraction of their energy into the radiation field.
At these energies, the electrons will mainly radiate at frequencies
$\nu \ge\nu_-$, where
\begin{equation}
\nu_- = (2.2\cdot 10^{13}\, {\rm Hz})\,\left(\frac{n_{\rm p}}{\rm 10^9\,cm^{-3}}\right)
  \left( \frac{B}{\rm G} \right)^{-1},
\label{eq-numinus}
\end{equation}
i.e. under optically thin conditions for the standard parameters
(see table \ref{tab-params}). So
\begin{eqnarray}
\qquad \nu_- &\gg& \nu_+ \nonumber \\
\Leftrightarrow \left(\frac{T}{\rm 10^4\,K}\right)^{3\over 4}
 &\gg& 0.14\ 
\sqrt{\frac{s}{\rm 10^{14}\,cm}}\,\sqrt{\frac{L}{\rm 10}}\,\left( \frac{B}{\rm G} \right) \label{eq-nu-nu+2}
\end{eqnarray}
which means that only a small fraction of the synchrotron emission
will be absorbed
by the plasma, and thus Coulomb heating should always be more efficient
than heating by absorption. A careful evaluation of Eq.\ref{eq-fsync}
using a Taylor expansion in $\nu_+\ll \nu_c$ indicates
that the numerical factor on the RHS of Eq.\ref{eq-nu-nu+2} should be closer
to unity, but, on the other hand, that equation does not consider
the heating by energetic protons.
We may therefore conclude that synchrotron heating can be neglected,
if the condition (\ref{eq-nu-nu+2}) is met.

\subsection{The balance of heating and cooling}
\label{chap-tempevol}
The evolution of the temperature is closely tied to the
evolution of the heating and cooling rates. 
For a homogeneous interstellar medium the system will tend to attain
a quasi-steady state. Consequently, the heating rates
rise with time until they reach an asymptotic value. The temperature then
evolves until the cooling balances the heating.

The cooling rate is a strong function of the temperature, with a
huge jump at $10^4\,$K as the dominating feature at low temperatures
(see fig.\ref{fig-lambda}).
To reach higher temperatures, the heating rate must be large enough
to compensate the hydrogen cooling,
or the system will stay at $10^4\,$K, where the ionization fraction of hydrogen
may be low, which results in an even lower turn-over frequency, so that
the entire emission may be optical thin. Additionally,
it is likely that under the conditions prevailent in the jet plasma,
the temperature dependence of the ionization fraction is weaker than
in the LTE calculations of \citet{dmc72cool} and \citet{rcs76cool},
thus causing a slower rise of the cooling curve. Nevertheless,
the system will have a stable region around $10^4\,$K for a wide
range of heating rates.

When the heating rate becomes larger
than the maximum of the cooling function at $10^5\,$K, the temperature
will increase to more than $10^8\,$K, beyond which bremsstrahlung cooling
may again balance the heating. Extrapolating the cooling function to
temperature $T\gtrsim 10^8\,$K is problematic, though, for one quickly
arrives at relativistic temperatures, $kT\gtrsim m_{\rm e} c^2$, at which
most of our assumptions and formulae are no longer valid.
At these temperatures, it is especially required to include pair
production into our equations, which increases the particle number in the
background plasma while reducing the amount of energy available
for the individual particle, thus efficiently limiting the
temperature of the system.
In addition, the concept of a stable isotropic Maxwell-Boltzmann
distribution is not well justified in these regions \citep{s82therm}. 

\begin{figure}
\resizebox{\hsize}{!}{\includegraphics{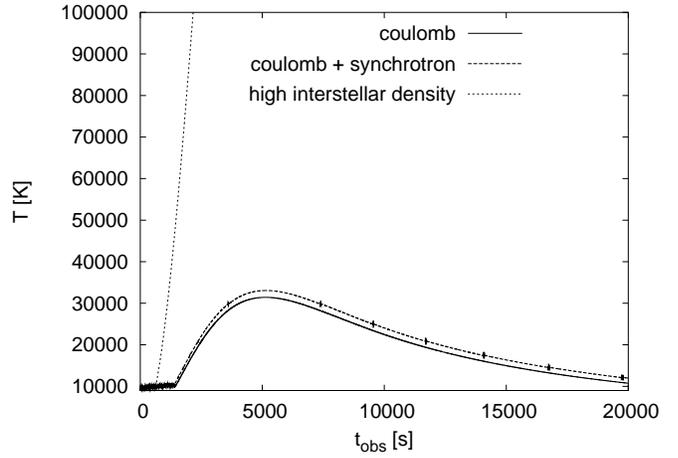}}
\caption{The temperature as a function of time for the standard
parameters as well as $n_{\rm i}^\ast=0.8\;\hbox{cm}^{-3}$, for which
the temperature becomes larger than $10^9\,$K,
where our model is invalid.}
\label{fig-temptime}
\end{figure}

For the standard parameters (see table \ref{tab-params}),
this overheating
does not happen (see fig. \ref{fig-temptime}), while
for $n_i^\ast=0.8\;\hbox{cm}^{-3}$ the heating processes are too
strong to keep the temperature non-relativistic. The Fig. also
demonstrates that the synchrotron contribution to the
heating rate is quite small; numerical evaluations show that
this process contributes about 10 \% of the Coulomb heating rate,
which is in good agreement with our estimate in the preceding
section.
Finally, the asymptotic cooling is a direct result of the
deceleration of the system, which decreases the number of
swept up particles (eq. \ref{eq-inje}).
At $t=0$, when there are no non-thermal particles present, the
heating rates equal zero, and the system usually cools down to the stable
point at $10^4\,$K. It is possible to prevent this with a careful
choice of initial parameters, which results in stable temperatures
of $T=10^8\,$K.

\section{Results}
\label{chap-parameter}
\subsection{Modifying the initial parameters}
In the last sections we have seen that the temperature of the background
plasma can be used to estimate the stability of our model.
For this reason, we
have made a parameter study to approximate the regions in which we
get reliable results. We have discovered four different regions:
cold ($T\leq 10^4\,$K), warm ($10^4\,$K$\leq T \leq 10^5\,$K), hot
($10^8\,$K$\leq T \leq 10^9\,$K) and forbidden ($T>10^9\,$K).
The region between $10^5\,$K and $10^8\,$K is unstable, which is a result of
the specific form of the cooling function. The region between $10^8$ and
$10^9\,$K is stable only for very specific initial parameters,
as outlined in the last section.

\begin{figure}
\resizebox{\hsize}{!}{\includegraphics{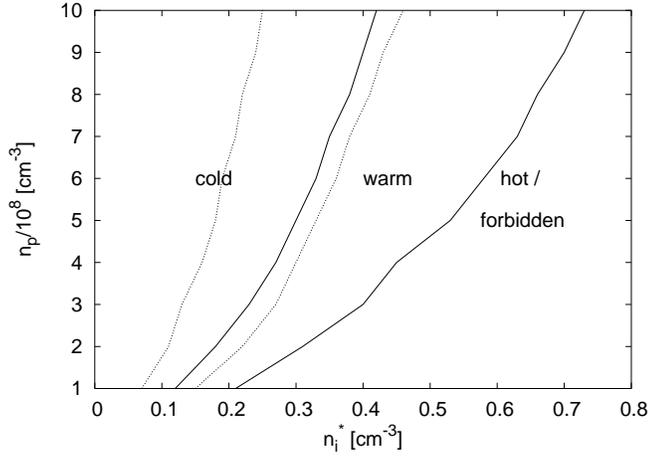}}
\caption{The three qualitatively different temperature regions as a
function of the particle densities and the standard parameters
($T_0 = 10^4\,{\rm K}$).
The dotted lines are for $\Gamma_0 = 600$.}
\label{fig-paramn}
\end{figure}

The particle number densities $n_{\rm p}$ and $n_{\rm i}^\ast$ strongly dominate
the system; the different regions in the $n_{\rm p}-n_{\rm i}^\ast$-plane for the
standard parameters (see table \ref{tab-params}) are displayed in Fig.\ref{fig-paramn}.
Another parameter which determines the temperature evolution
is the size of the system. It turns out that on account of the
particle number sweep-up rate (eq.\ref{eq-inje}), which is
proportional to the sweep-up surface divided by the total volume
of the system, only one of the size
parameters really contributes to this point, which can be
identified as the average optical path for a generic
system. For the
specific disk geometry, this parameter is identical to the thickness $d$;
then for the standard parameters the boundary points are
\begin{eqnarray}
d_{cold-warm} &\approx& 2.1 \cdot 10^{13}\  \hbox{cm}\\
d_{warm-hot\ } &\approx& 4.1 \cdot 10^{13}\  \hbox{cm}
\end{eqnarray}
Since the secondary electrons dominate the heating process (see Sect.
\ref{chap-heat-el}), and their production rate can not be treated
analytically, these results can not be derived from Fig. \ref{fig-paramn}.

The initial Lorentz factor $\Gamma(t=0)$ modifies the interstellar matter
density in the jet frame, $n_{\rm i}=\Gamma(t) n_{\rm i}^\ast$. This in turn modifies
the number of particles in the system and therefore the heating rate by
elastic scattering. However, it does not modify the \emph{form} of the
boundaries of temperature regions in the $n_{\rm p}-n_{\rm i}^\ast$-plane, but only their
\emph{location}. To demonstrate this, in fig. \ref{fig-paramn}
we also display the temperature regions for $\Gamma_0 = 600$.

The observer angle $\theta^{\rm obs}$ only modifies the observed time required
for the system to evolve, so the asymptotic behaviour
remains unchanged.
The magnetic field strength $B$ has practically no influence
on the temperature, since the synchrotron heating process
is negligible.

As mentioned earlier, by modifying the initial temperature $T_0$
together with the interstellar matter density $n_{\rm i}^\ast$,
we are able to establish stable 'hot' temperatures. In most cases,
however, the initial temperature will not modify the system.

From this we conclude that our model is stable only on a rather
small region in the initial parameter space. In this case, there
are three asymptotically stable temperature regions, and it
may be possible to fix some of these parameters by observations.
In all other cases, the initial assumption of a non-relativistic
thermal component in the jets breaks down, although the sweep-up
mechanism itself remains valid.

\subsection{Time-dependent variations of free parameters}
In a real situation the extragalactical matter density
$n_{\rm i}^\ast$ is not constant over long timescales and distances.
The plasma in the jet may cross dense clouds of hydrogen
or enter a local fluctuation. For this reason we have investigated
the temperature evolution under
some simple modifications of our sweep-up rate.

First, we have considered a simple step in the ambient
matter density,
\begin{eqnarray}
\dot{N}_{cld} &=& \dot{N} (1+a \Theta(t_1-t)).
\label{eq-cloud}
\end{eqnarray}
Such a situation may possibly be found when the jet leaves the
immediate host galaxy and enter the intergalactic medium.
In this case it is slightly easier to produce stable temperatures
at $10^8\,$K, although this situation will still be quite rare.
Generally, this modification simply produces a break
in the temperature evolution.

Secondly, we have investigated the effect of local density fluctuations.
We have studied a situation in which the system periodically passes
through a dense cloud of material, that is embedded in low-density gas,
where for a single period $\tau$ the sweep-up rate is
\begin{eqnarray}
\dot{N}_{peak}(t) &=&
 \,\left\{
  \begin{array}{rl}
  a&\mbox{for }t < c\\
  b&\mbox{for }c < t < \tau
  \end{array} \right.
\label{eq-peak}
\end{eqnarray}
with $a \gg b$ and $c$ the fraction
of the period where the heating is strong. We have normalized these
parameters in a way that the interstellar matter density averaged
over a period is $n_{\rm i}^\ast$.
Under this periodically peaked scheme, it is again
easier to keep the system stable at $10^8\,$K without causing
overheating as described in chapter \ref{chap-tempevol}.

\subsection{Light curves and correlations}
\begin{figure}
\resizebox{\hsize}{!}{\includegraphics{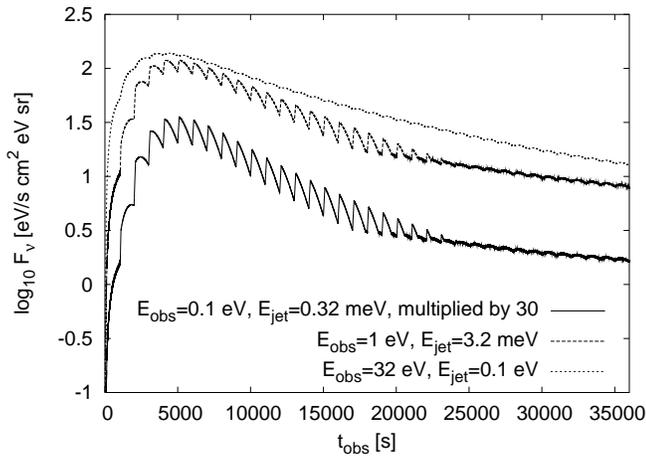}}
\caption{Light curves for three different energies in the observer frame.
Here we have adopted the periodic peak injection introduced
in eq. \ref{eq-peak} with the parameters $a/b=50$, a period of
$\tau = 1000\,{\rm s}$ observed time and $c=100\,{\rm s}$, i. e. 10\% of the period time.
Significant variations occur exclusively in the optical thick regime; 
the redshift of the source has been set to $z=0.5$.
}
\label{fig-light}
\end{figure}

We have then calculated light curves for several different energies,
which we present in Fig. \ref{fig-light} for the periodically peaked
injection and the standard parameters. Unlike the other photon spectra
presented in this report, these curves are taken in the observer frame,
in which, on account of the deceleration of the jet,
the frequency corresponding to a fixed jet-frame frequency varies with time,
although the modification usually is small.

We see that the variability is visible, but that the average
behaviour of the system remains unchanged. Additionally,
the variability only affects the optical thick region, which
is a result of temperature variations. In the optical thick region,
we have $I \propto j/\alpha$ (eq. \ref{eq-inu1b}) and
$\alpha \approx \alpha_{\rm thermal}$, which in turn causes $F_\nu$ to depend
on the temperature. Since the heating rate is proportional to
the total number of energetic particles, a sudden increase in
this number causes a similar modification of the temperature
of the jet plasma, which again modifies the optical thick emission.
Because of $j \approx j_{\rm sy}$, the optical thin emission
(see eq. \ref{eq-inu1a}) does not demonstrate a similar behaviour.
For this reason our model is able to reproduce different variability
in different freqnency bands.

Next, we have calculated correlation coefficients for this injection
profile, using light curves in the observer frame between the
infrared, optical and ultraviolet, from the plasma frequency up
to a few 10 eV, with small and large distances between the considered
frequencies.

It turns out that the lightcurves at different frequencies are almost
perfectly correlated ($> 0.9$), even if we compare 
free-free-dominated with synchrotron-dominated parts of the spectrum.
To understand
this behaviour, we first note that the electron distribution
is dominated by the diffusive escape process (eq. \ref{eq-tesc}),
while the continuous losses are too inefficient. For high-energy
electrons the particle spectrum is exclusively determined by
the Lorentz factor of the jet, $\Gamma$.
For this reason all light curves produced by the electrons are highly
correlated with each other, even between the free-free and synchrotron
processes, since the free-free emission depends on the temperature, which
again is determined by the high-energy electrons, similar to the
synchrotron emission. Given that the optically thin synchrotron emission
shows only a very weak response to the variations in the injection rate,
most of the correlation signal is presumably caused by the common secular
trend in all light curves that is caused by the deceleration of the jet.

\section{Possible corrections}
In this section we briefly discuss possible further aspects of
the problems investigated in this paper.

\subsection{The electrostatic instability}
\label{sec-esi}
Recently, \citet{p02esi} have investigated the electrostatic
instability in a relativistic beam of electrons and protons
and its impact on our model. Their conclusion was that
the sweep-up spectrum effectively gets modified from a peak structure
at $\gamma=\Gamma$ to a plateau distribution up to $\gamma=\Gamma$,
while keeping the number of collected particles and the low-energy
photon spectrum unchanged.
The energy lost in this process may contribute to the heating
of the background plasma, but there is no reliable estimate available of
the exact amount of energy transferred. Observations hint that
only a tiny fraction of the total energy lost by this mechanism
can ever contribute to the heating.

Since this process in principle involves a huge amount of energy,
it might dramatically change the evolution of the temperature. However,
\citet{p02esi} have only considered the initial situation of a cold 
electron-proton beam penetrating cold thermal plasma. The presence of isotropic
energetic particles should strongly reduce the growth rate of electrostatic 
turbulence, so that at least in a quasi-steady state heating by
damping of electrostatic waves should be much less severe than
suggested by the asymptotic energy loss of the incoming particle beam.

\subsection{The initial conditions}
Currently, our initial conditions assume that there are no non-thermal
particles present. Although the mechanism responsible for creation
and stability of the jets is not completely understood
\citep{urry95,bbr84theory}, it is likely that there is no time zero,
at which a plasma cloud is relativistically expelled without
containing energetic particles. Because of this, the start-up phase
visible in our results would be
a consequence of our particular treatment of the problem, and not
a physical phenomenon.

In section \ref{chap-parameter} we have found that only a few
of our parameters significantly
modify the asymptotic behaviour. The average system size only
affects the observed photon spectra around the turn-over frequency,
where $\tau = 1$, and the particulars of the geometry are unknown anyway.
The initial particle distributions won't modify the
asymptotic temperature evolution either, because the presence of a moderate
amount of non-thermal particles only modifies the time required to reach
a balance of gains and losses. Variations of the matter densities
have already been investigated in section \ref{chap-parameter}.
All other initial parameters do not modify the general behaviour of
the system.

\section{Summary and conclusions}
\label{chap-conclusions}
In this paper we have investigated the evolution of thermal
plasma in AGN jets and its impact on the
optical-to-infrared photon emission. In 
matter-dominated jets sizable quantities of gas should exist, in particular
in the compression zones near the collision fronts that are an implicite ingredient of
Fermi-type particle acceleration scenarios.
We conduct our study in the framework of the
channeled outflow model of \citep{ps00blast}, who have studied the kinetic
relaxation of particles that have traversed a parallel collision front
from the
upstream to the downstream region. This pick-up process occurs due to 
scattering off low-wavenumber plasma waves that the picked-up 
particles generate themselves through streaming instabilities.
The calculations, thus, describe the first half-cycle of the standard shock
acceleration process for relativistic, collisionless flows.
If the outflow plasma propagating through the jet is dense,
the downstream region provides a target for the 
ultra-relativistic protons to interact with and produce radiation
through various leptonic and hadronic emission channels.

Non-relativistic thermal gas in AGN jets
noticably contributes to the optical depth at radio to infrared frequencies, and much less
to the emission, with an efficiency that is strongly temperature-dependent.
Assuming that this plasma is in a thermal equilibrum, we have calculated the temperature
evolution resulting from the competition of radiative cooling and heating by Coulomb processes
and absorption of non-thermal emission. 

Similar to the well-known results for the structure of the interstellar medium in Galaxies,
we find that the stable regimes exist for temperatures
between $10^4\,$K and $10^5\,$K, and around $10^8\,$K.
Below $10^4\,$K the ionization fraction will be small, and the optical depth is modified.
Above $10^8\,$K, the thermal particles reach relativistic velocities,
for which our model begins to break down. 

In the model of \citet{ps00blast}, short-time variability
at low energies arises on account of 
density fluctuations in the upstream medium.
Consequently the plasma temperature will vary, and so will
the thermal absorption coefficient. The observable flux of
low-frequency emission is thus modulated by the temperature evolution of the
thermal gas, and it can therefore display very complicated variability
behaviour. For simple density profiles of the interstellar gas in AGN host galaxies,
we have calculated the temperature response of the thermal gas in the jet, and have
then derived light curves at different frequencies.
For sufficiently long observing times,
all of these light curves turn out to be strongly correlated with
each other, independent of the proton injection scheme, only
on account of the deceleration of the jet.

\begin{acknowledgements}
Partial support by the Bundesministerium 
f\"ur Bildung und Forschung through DESY,
grant 05 CH1PCA6, is gratefully acknowledged.
\end{acknowledgements}

%
%
\begin{appendix}
\section{Solving the radiation transport equation with
the exact cylinder geometry}
\label{app-geometry}
The most general solution to the radiation transport equation
(\ref{eq-radtrans}) is
\begin{eqnarray}
I_\nu &=& S_\nu \int dA (1-\exp(-\tau(x, \Omega))
\label{eq-intensityx}
\end{eqnarray}
where the optical thickness depends on the emission point $x$
and the emission direction $\Omega$. The optical thickness
needs to be calculated with respect to the geometry of the system.
In the case of the cylinder geometry this expression can be evaluated,
with several one-dimensional integrals remaining which must be
treated numerically.

\begin{figure}
\resizebox{\hsize}{!}{\includegraphics{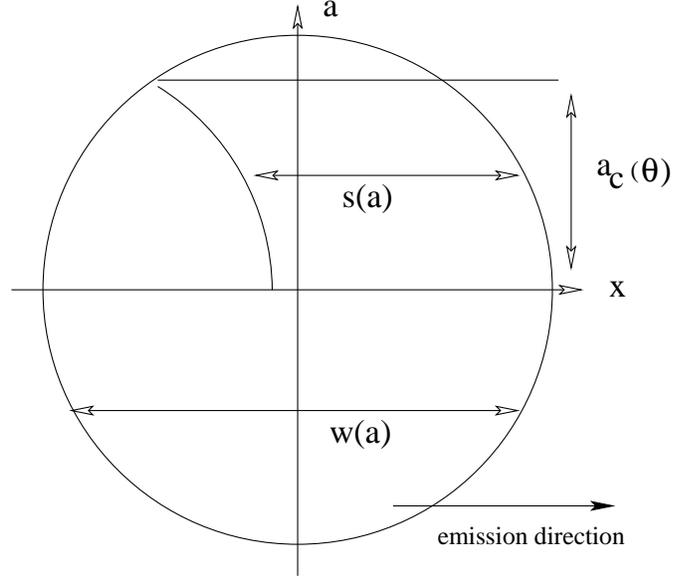}}
\caption{Definitons used in the text to calculate the integral
over the emitting surface. Top view.}
\label{fig-geometry}
\end{figure}

To solve equation \ref{eq-intensityx}, we treat the emitting surface
in \emph{Cartesian} coordinates. (If we used polar coordinates, we
would end up with two-dimensional numerical integrals.)
We define our coordinate system as presented in fig.\ref{fig-geometry}.
Then the problem reduces to an integral over all impact parameters $a$
and an integral over all 'slices' $x=-w(a)..w(a)$.
We will evaluate the integrals over the 'slices' first.

\begin{figure}
\resizebox{\hsize}{!}{\includegraphics{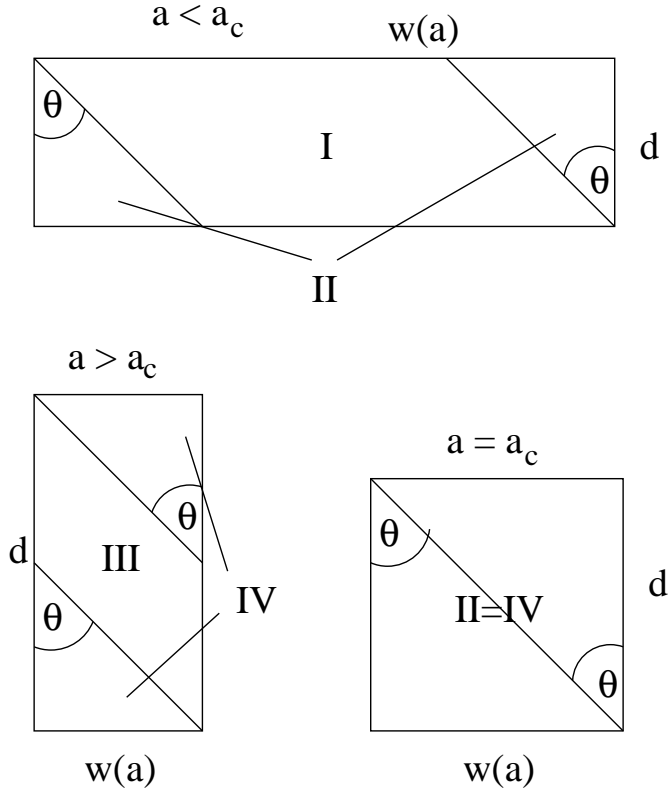}}
\caption{Possible configurations of the integral over $x$ as a
function of the other parameters.}
\label{fig-slices}
\end{figure}

It is useful to define the following quantities, related to the
selected coordinate system (see fig. \ref{fig-geometry}), which
follow from elemental trigonometry.
\begin{eqnarray}
\theta_c &=& \arctan \frac{2r}{d} 
\label{eq-thetac} \\
a_c(\theta) &=& \sqrt{r^2-d^2/4 \tan^2\theta}
\label{eq-ac} \\
w(a) &=& 2\sqrt{r^2 - a^2}
\label{eq-width} \\
s(\theta) &=& d \tan \theta
\label{eq-svona}
\end{eqnarray}
The angle $\theta_{\rm c}$ is the critical emission angle, above which some
contributions to the integral vanish. $a_c(\theta)$ is the critical
impact parameter where the mathematical form of the slices
changes (see fig. \ref{fig-slices} for a sketch of this).
Above $\theta=\theta_c{\rm c}$ this expression no longer makes any sense.
$w(a)$ is the
'width' of the system for an impact parameter $a$ and determines
the integration limits for the variable $x$. Finally $s(\theta)$ is
used to divide the slice integral in the two
different regions called $I$ and $II$ in fig. \ref{fig-slices}.
With these conventions the integral splits in
\begin{eqnarray}
I_\nu &=& I_{I+II} + I_{III+IV} \\
 &=& 2(\int_{0}^{a_c} S_1(\theta) da + \int_{a_c}^r S_2(\theta) da)
\label{eq-islices}
\end{eqnarray}
where $S_i$ are the two types of slices that are possible in the system
(see fig. \ref{fig-slices}).

\begin{figure}
\resizebox{\hsize}{!}{\includegraphics{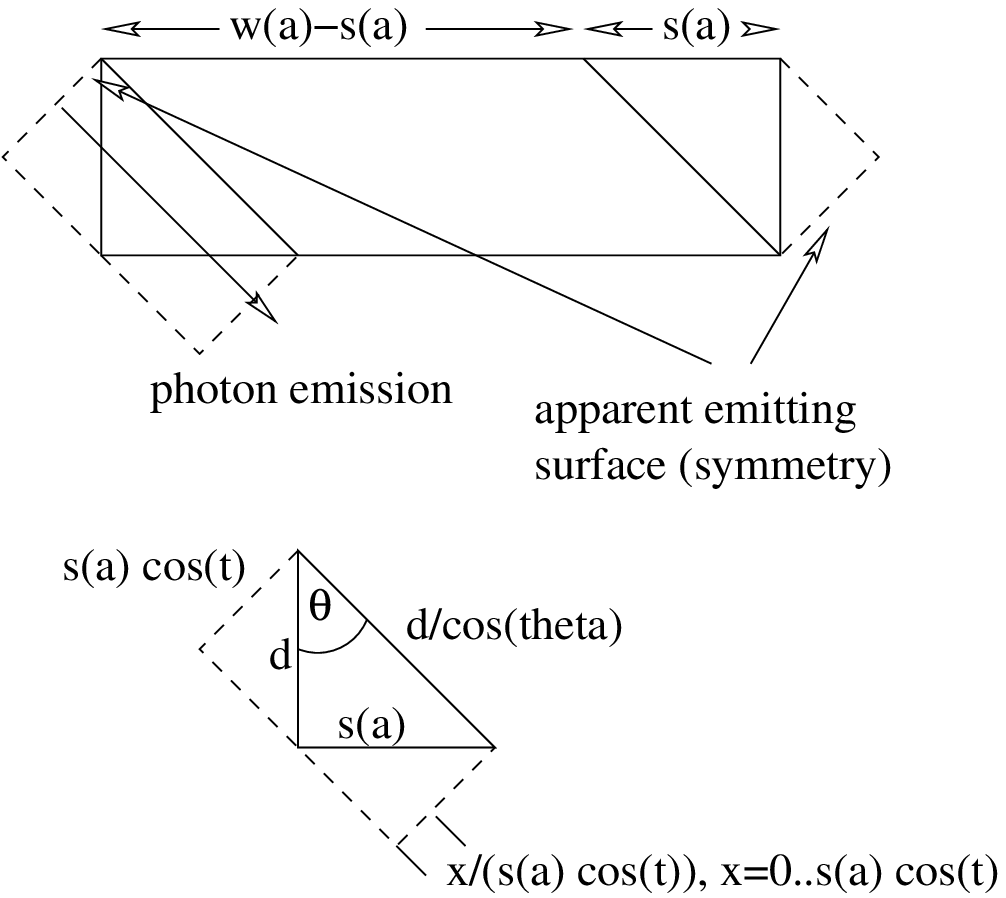}}
\caption{Detailed view of the geometry in the x-direction
for small 'impact parameters' $a$.}
\label{fig-slice12}
\end{figure}

\begin{figure}
\resizebox{\hsize}{!}{\includegraphics{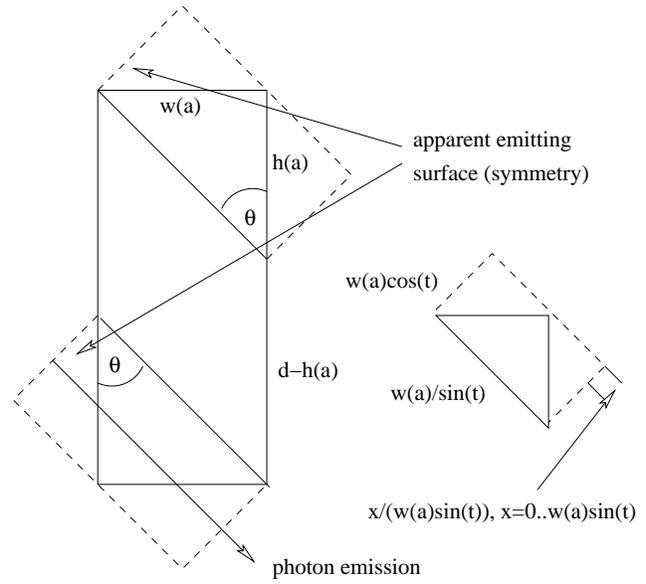}}
\caption{Detailed view of the geometry in the x-direction
for large 'impact parameters' $a$.}
\label{fig-slice34}
\end{figure}

The first type of slice is the inner part of the disk (for small $a$),
where some of the emitted photons (I) see a constant path length, where
only the border region (II) is modified. Then the integral over
$S_1$ can be solved exactly, and we get
\begin{eqnarray}
S_1 &=& (w(a)-s(\theta)) \cos \theta (1-\exp(-\tau_d/\cos \theta )) \nonumber \\
 && + 2 \int_0^{s(\theta) \cos \theta}
   (1-\exp(\alpha\frac{d}{\cos\theta}\frac{x}{s(\theta) \cos\theta})) dx \\
 &=& (1-\exp(-\tau_d/\cos \theta ))\cdot \nonumber \\
 && ((w(a)-s(\theta)) \cos \theta - 2 \frac{\sin\theta \cos\theta}{\alpha})
  \nonumber \\
 && + 2 s(\theta) \cos\theta.
\label{eq-s1}
\end{eqnarray}
To solve the integral over $x$ we have used the parameterization
$\tau(x) = \tau_{\rm max} \cdot x/x_{\rm max}$, with $0 \leq x \leq x_{\rm max}$
(see fig.\ref{fig-slice12}).
Inserting this expression in \ref{eq-islices} and integrating over
$a$ gives us
\begin{eqnarray}
I_1 &=& 2(1-\exp(-\tau_d/\cos \theta ))\cdot \nonumber \\
 && (\cos \theta (a_c w(a_c)/2+r^2 \arcsin{a_c/r}) -a_c d \sin \theta
  \nonumber \\
 && -2 \frac{a_c \sin\theta \cos\theta}{\alpha}) \nonumber \\
 && + 4 a_c d \sin\theta \equiv I_{I+II}
\label{eq-i1}
\end{eqnarray}
which is the exact solution of the radiation transport equation
for the region defined by $a<a_{\rm c}$.

The second type of slice, $S_2$ contains expressions that can
not be solved analytically. The general structure of the integrals
involved here is very similar to the $S_1$ contributions,
\begin{eqnarray}
S_2 &=& 2 \int_0^{w(a)\cos\theta} 
 (1-\exp(-\alpha \frac{w(a)}{\sin\theta}\frac{x}{w(a)\cos\theta})) dx
  \nonumber \\
 && + (d-h(a))\sin\theta (1-\exp(-\tau_{w(a)}/\sin\theta )) \\
 &=& 2 w(a) \cos\theta \nonumber \\
 && + (1-\exp(-\tau_{w(a)}/\sin\theta )) \cdot \nonumber \\
 && ((d-h(a))\sin\theta - 2\frac{\sin\theta \cos\theta}{\alpha}).
\label{eq-s2}
\end{eqnarray}
Inserting this in eq. \ref{eq-islices} results in non-analytical
integrals,
\begin{eqnarray}
I_2 &=&
 \cos\theta ( 2 \arccos \frac{a_c}{r} -a_c w(a_c))
  \nonumber \\
 && + 2 \sin\theta (r-a_c)(d-\frac{2 \cos\theta}{\alpha})
  \nonumber \\
 && + 2 \cos\theta \int_{a_c}^r w(a)
  \exp(-\frac{\tau_{w(a)}}{\sin(\theta)})da)
  \\
 && - 2 \sin\theta (d - 2\frac{\cos\theta}{\alpha})
   \int_{a_c}^r \exp(-\frac{\tau_{w(a)}}{\sin(\theta)})da \equiv I_{III+IV}
 \nonumber
\label{eq-i2}
\end{eqnarray}
The remaining integrals are of the form $\int \exp(-\sqrt{1-a^2}) f(a) da$,
which can only be solved in very special cases, such as $f(a)=a$.
However, these integrals can be evaluated numerically in
a fast and reliable way with the use of simple quadrature formulae
of low order (see, e.g. \citet{nrfor}).

Finally, for $\theta > \theta_{\rm c}$ the entire contribution
$I_1$ vanishes, since in this case
all slices are of the type '2',
where all photons emitted from the top surface will
leave the system through the side.
So the total photon intensity emitted by a plasma in the form of
a thin disk is
\begin{eqnarray}
I(\theta<\theta_c) &=& I_1(a_c(\theta))+I_2(a_c(\theta)) \\
I(\theta>\theta_c) &=& I_2(a_c=0).
\end{eqnarray}

\section{Solving the radiation transport equation with
the exact square box geometry}
\label{app-box}
The exact solution of the radiation transport equation takes a
much simpler expression if one uses a box geometry.
In fact, for this geometry all integrals are of the same type as
the region called '1' in the last appendix, which does not involve any
kind of numerical integrations.
To keep the volume of the system constant, we use $a=\sqrt{\pi}r$
for the quadratic top and bottom surfaces and get
\begin{eqnarray}
I_{box}(\theta < \theta_c) &=& a(1-\exp(\tau_d/\cos{\theta}))\cdot
  \nonumber \\
 && (d \sin\theta - a\cos\theta - 2\frac{\sin\theta \cos\theta}{\alpha})
  \nonumber  \\
 && + 2ad\sin\theta \\
I_{box}(\theta > \theta_c) &=& a(1-\exp(\tau_a/\sin{\theta}))\cdot
  \nonumber \\
 && (a \cos\theta - d \sin\theta - 2\frac{\sin\theta \cos\theta}{\alpha})
  \nonumber  \\
 && + 2a^2\cos\theta.
\end{eqnarray}
\end{appendix}

\bibliographystyle{aa}

\begin{thebibliography}{35}
\expandafter\ifx\csname natexlab\endcsname\relax\def\natexlab#1{#1}\fi

\bibitem[{Begelman {et~al.}(1984)Begelman, Blandford, \& Rees}]{bbr84theory}
Begelman, M.~C., Blandford, R.~D., \& Rees, M.~J. 1984, Rev. Mod. Phys., 56,
  255

\bibitem[{Butler \& Buckingham(1962)}]{bb62}
Butler, S. \& Buckingham, M. 1962, Phys. Rev., 126, 1

\bibitem[{Capella {et~al.}(1994)Capella, Sukhatme, Tan, \&
  Tran}]{cap94dualparton}
Capella, A., Sukhatme, U., Tan, C.-I., \& Tran, T. V.~J. 1994, Phys. Rep, 236,
  227

\bibitem[{Corbel {et~al.}(2002)Corbel, Fender, Tzioumis, {et~al.}}]{corbel02}
Corbel, S., Fender, R., Tzioumis, A., {et~al.} 2002, Science, 298, 196

\bibitem[{Crusius \& Schlickeiser(1988)}]{cs88rad}
Crusius, A. \& Schlickeiser, R. 1988, A\&A, 196, 327

\bibitem[{Dalgarno \& McCray(1972)}]{dmc72cool}
Dalgarno, A. \& McCray, R.~A. 1972, Ann. Rev. A\&A, 10, 375

\bibitem[{Dermer \& Gehrels(1995)}]{der95obs}
Dermer, C.~D. \& Gehrels, N. 1995, Astrophys. J., 441, 270

\bibitem[{Elliot \& Shapiro(1974)}]{es74relation}
Elliot, J.~D. \& Shapiro, S.~L. 1974, Astrophys. J., 192, L3

\bibitem[{Engel {et~al.}(1997)Engel, Ranft, \& Roesler}]{dtunuc97}
Engel, R., Ranft, J., \& Roesler, S. 1997, Phys. Rev. D, 55, 6957

\bibitem[{Ferrari {et~al.}(1996a)Ferrari, Sala, Ranft, \&
  Roesler}]{ferr96dtunuc}
Ferrari, A., Sala, P.~R., Ranft, J., \& Roesler, S. 1996a, Z. Phys. C, 70, 413

\bibitem[{Ferrari {et~al.}(1996b)Ferrari, Sala, Ranft, \& Roesler}]{Fer96b}
---. 1996b, Z. Phys. C, 71, 75

\bibitem[{Georganopoulos \& Kazanas(2003)}]{gk03}
Georganopoulos, M. \& Kazanas, D. 2003, Astrophys. J., 594, L27

\bibitem[{Haug(1988)}]{h88coulomb}
Haug, E. 1988, A\&A, 191, 181

\bibitem[{Hayakawa(1969)}]{haya69}
Hayakawa, S. 1969, Cosmic Ray Physics (Wiley-Interscience)

\bibitem[{Homan {et~al.}(2003)Homan, Lister, Kellermann, {et~al.}}]{hom03}
Homan, D.~C., Lister, M.~L., Kellermann, K.~I., {et~al.} 2003, Astrophys. J.,
  589, L9

\bibitem[{Homan {et~al.}(2002)Homan, Wardle, Cheung, Roberts, \&
  Attridge}]{hom02}
Homan, D.~C., Wardle, J. F.~C., Cheung, C.~C., Roberts, D.~H., \& Attridge,
  J.~M. 2002, Astrophys. J., 4580, 742

\bibitem[{Jauch \& Rohrlich(1976)}]{jr76book}
Jauch, J.~M. \& Rohrlich, F. 1976, The Theory of Photons and Electrons
  (Springer-Verlag)

\bibitem[{Konopelko {et~al.}(2003)Konopelko, Mastichiadis, Kirk, Jager, \&
  Stecker}]{kono03}
Konopelko, A., Mastichiadis, A., Kirk, J., Jager, O.~D., \& Stecker, F. 2003,
  ApJ, 597, 851

\bibitem[{M\"ohring \& Ranft(1991)}]{mr91dtunuc}
M\"ohring, H.-J. \& Ranft, J. 1991, Z. Phys. C, 52, 643

\bibitem[{M\"ohring {et~al.}(1993)M\"ohring, Ranft, Merino, \& Pajares}]{Moe93}
M\"ohring, H.-J., Ranft, J., Merino, C., \& Pajares, C. 1993, Phys. Rev. D, 47,
  4142

\bibitem[{Mukherjee {et~al.}(1997)Mukherjee, Bertsch, Bloom, {et~al.}}]{muk97}
Mukherjee, R., Bertsch, D.~L., Bloom, S.~D., {et~al.} 1997, Astrophys. J., 490,
  116

\bibitem[{Pacholczyk(1970)}]{p80book}
Pacholczyk, A.~G. 1970, Radio Astrophysics (Freeman, San Francisco, CA)

\bibitem[{Pohl {et~al.}(2002)Pohl, Lerche, \& Schlickeiser}]{p02esi}
Pohl, M., Lerche, I., \& Schlickeiser, R. 2002, A\&A, 383, 309

\bibitem[{Pohl \& Schlickeiser(2000)}]{ps00blast}
Pohl, M. \& Schlickeiser, R. 2000, A\&A, 354, 395

\bibitem[{Press {et~al.}(1986-2001)Press, Teukolsky, Vetterling, \&
  Flannery}]{nrfor}
Press, W.~H., Teukolsky, W.~A., Vetterling, W.~T., \& Flannery, B.~P.
  1986-2001, Numerical Recipes in Fortran (Cambridge University Press)

\bibitem[{Ranft {et~al.}(1994)Ranft, Capella, \& Tr\^an}]{ranft94dtunuc}
Ranft, J., Capella, A., \& Tr\^an, T. V.~J. 1994, Phys. Lett B, 320, 346

\bibitem[{Ranft \& Roesler(1994)}]{RR94}
Ranft, J. \& Roesler, S. 1994, Z. Phys. C, 62, 329

\bibitem[{Raymond {et~al.}(1976)Raymond, Cox, \& Smith}]{rcs76cool}
Raymond, J.~C., Cox, D.~P., \& Smith, B.~W. 1976, Ap. J., 204, 290

\bibitem[{Roesler {et~al.}(1998)Roesler, Engel, \& Ranft}]{RER98}
Roesler, S., Engel, R., \& Ranft, J. 1998, Phys. Rev. D, 57, 2889

\bibitem[{Rybicki \& Lightman(1985)}]{rl85radiation}
Rybicki, R.~B. \& Lightman, A.~P. 1985, Radiative processes in astrophysics
  (Jon Wiley \& Sons)

\bibitem[{Schlickeiser(2002)}]{rsch02book}
Schlickeiser, R. 2002, Cosmic Ray Astrophysics (Springer-Verlag)

\bibitem[{Stepney(1982)}]{s82therm}
Stepney, S. 1982, Mon. Not. R. Astr. Soc., 202, 467

\bibitem[{Urry \& Padovani(1995)}]{urry95}
Urry, C.~M. \& Padovani, P. 1995, PASP, 107, 803

\bibitem[{Vainio {et~al.}(2004)Vainio, Pohl, \& Schlickeiser}]{vps03}
Vainio, R., Pohl, M., \& Schlickeiser, R. 2004, A\&A, 414, 463

\bibitem[{Vermeulen \& Cohen(1994)}]{vc94}
Vermeulen, R.~C. \& Cohen, M.~H. 1994, Astrophys. J., 430, 467

\end{thebibliography}

\end{document}